\documentclass[a4paper,12pt]{article}

\usepackage[dvips]{graphicx}
\usepackage{epsfig}

\begin{document}

\author{C. Barrab\`es\thanks{E-mail : barrabes@celfi.phys.univ-tours.fr}\\
\small Laboratoire de Math\'ematiques et Physique Th\'eorique,\\
\small CNRS/UMR 7350, F\'ed\'eration Denis Poisson CNRS/FR2964, \\
\small Universit\'e F. Rabelais, 37200 TOURS,
France
\\\small and \\P. A. Hogan\thanks{E-mail : peter.hogan@ucd.ie}\\
\small School of Physics,\\ \small  University College Dublin, Belfield, Dublin 4, Ireland}

\title{Generating a Cosmological Constant with Gravitational Waves}
\date{}
\maketitle

\begin{abstract}A technique is given to derive the well known Bell--Szekeres solution of the Einstein--Maxwell vacuum field equations describing the space--time and the Maxwell field 
following the head--on collision of two homogeneous, plane, electromagnetic shock waves. The analogue of this technique is then utilized to construct the space--time model of the gravitational field 
following the head--on collision of two homogeneous, plane, gravitational shock waves. The latter collision, which is followed by a pair of impulsive gravitational waves and a pair of light like 
shells traveling away from each other,  provides a mechanism for generating a cosmological constant which may be important 
in the theoretical description of dark energy. 
\end{abstract}
\thispagestyle{empty}
\newpage

\section{Introduction}\indent
This paper is concerned with exploring the strong analogy between the collision of electromagnetic shock waves and the collision of gravitational shock waves and with demonstrating that the 
latter is a mechanism for generating a cosmological constant, which may be important in the theoretical description of dark energy \cite{Pe}. It is important in this context to note that the appearance of a cosmological constant term on the left--hand side of the Einstein field equations is 
equivalent to the appearance of an energy--momentum--stress tensor for a perfect fluid with an equation of state expressing the vanishing of the sum of the matter proper density and the 
isotropic pressure. Thus for the case of colliding gravitational shock waves the space--time consists of an anti--collision region which is a vacuum and a post--collision region which is a 
non--vacuum in this sense. Vacuum and non--vacuum regions of space--time are familiar from solving the field equations for so--called interior and exterior solutions. Immediately following 
the collisions considered here a pair of impulsive gravitational waves is formed, moving away from each other, for the case of colliding electromagnetic shock waves (which is a well--known 
phenomenon \cite{BS}) while following the collision of two gravitational shock waves a pair of impulsive gravitational waves \emph{and} a pair of light--like shells, moving away from each other, are formed.

The collision of plane electromagnetic shock waves with a Heaviside step function profile has been solved in Einstein--Maxwell theory many years ago by Bell and Szekeres \cite{BS}. The collision 
is head--on and the waves are homogeneous. Such waves are described by a solution of the vacuum Einstein--Maxwell field equations starting with the line element
\begin{equation}\label{1.1}
ds^2=-\cos^2au_+(dx^2+dy^2)+2\,du\,dv\ ,\end{equation}where $a$ is a real constant and $u_+=u\vartheta(u)$ with $\vartheta(u)$ the Heaviside step function which is equal to zero if $u<0$ and equal 
to unity if $u>0$. If we write the line element in terms of the basis 1--forms $\vartheta^1=\cos au_+\,dx, \vartheta^2=\cos au_+\,dy, \vartheta^3=dv$ and $\vartheta^4=du$ we have
\begin{equation}\label{1.2}
ds^2=-(\vartheta^1)^2-(\vartheta^2)^2+2\vartheta^3\,\vartheta^4=g_{ab}\vartheta^a\vartheta^b\ ,\end{equation}where the constants $g_{ab}=g_{ba}$ are the components of the metric tensor on the half null 
tetrad defined by the basis 1--forms. Tetrad indices (such as $a, b$ here) will be lowered and raised using $g_{ab}$ and $g^{ab}$ respectively, with $g^{ab}$ the components of the inverse of the matrix 
with entries $g_{ab}$ and thus $g^{ab}g_{bc}=\delta^a_c$. The components of the Ricci tensor on this tetrad are given by
\begin{equation}\label{1.3}
R_{ab}=-2a^2\vartheta(u)\delta^4_a\delta ^4_b\ .\end{equation}The Maxwell field is described by the 2--form
\begin{equation}\label{1.4}
F=\frac{1}{2}F_{ab}\vartheta^a\wedge\vartheta^b=a\,\vartheta(u)\vartheta^1\wedge\vartheta^4\ ,\end{equation}(or equivalently by the Newman--Penrose \cite{NP} component $\Phi_2=a\,\vartheta(u)$) and the corresponding 
electromagnetic energy--momentum tensor is
\begin{equation}\label{1.5}
E_{ab}=F_{ac}F^c{}_b-\frac{1}{4}g_{ab}\,F_{dc}F^{dc}=-a^2\vartheta(u)\delta^4_a\delta^4_b\ .\end{equation}Hence (\ref{1.3}) and (\ref{1.5}) demonstrate that the Einstein--Maxwell vacuum field equations
\begin{equation}\label{1.6}
R_{ab}=2\,E_{ab}\ ,\end{equation}are satisfied. The Hodge dual of the Maxwell 2--form (\ref{1.4}) is
\begin{equation}\label{1.7}
{}^*F=a\vartheta(u)\vartheta^2\wedge\vartheta^4\ ,\end{equation}which clearly satisfies the vacuum Maxwell field equations
\begin{equation}\label{1.8}
d{}^*F=0\ ,\end{equation}where $d$ denotes exterior differentiation. The Maxwell field (\ref{1.4}) is type N (the radiative type) in the Petrov classification with the vector field $\partial/\partial v$ as degenerate 
principal null direction and thus represents pure electromagnetic radiation. The profile of the wave is the Heaviside step function and so the wave is a shock wave. The coefficient of the step function in 
(\ref{1.4}) is a constant (in particular independent of the coordinates $x, y$) and thus the waves are homogeneous. The histories of the wave fronts in space--time are the null hyperplanes $u={\rm constant}$ 
which are generated by the null, geodesic, shear--free (and, trivially, twist--free) integral curves of the vector field $\partial/\partial v$. In the space--time with line--element (\ref{1.1}) the 
vector field $\partial/\partial u$ is null and generates the null hyperplanes $v={\rm constant}$.  The latter can be the histories of the wave fronts of electromagnetic shock waves traveling in the opposite direction to 
those with histories $u={\rm constant}$ and are given by the solution of the Einstein--Maxwell vacuum field equations consisting of the line element
\begin{equation}\label{1.9}
ds^2=-\cos^2bv_+(dx^2+dy^2)+2\,du\,dv\ ,\end{equation}where $b$ is a real constant and $v_+=v\vartheta(v)$, and the Maxwell field is given by the 2--form
\begin{equation}\label{1.10}
F=\frac{1}{2}F_{ab}\vartheta^a\wedge\vartheta^b=b\,\vartheta(v)\vartheta^3\wedge\vartheta^1\ ,\end{equation}or equivalently by the Newman--Penrose component $\Phi_0=b\,\vartheta(v)$. Bell and Szekeres 
considered the collision of two such families of electromagnetic shock waves, in the sense that if the space--time and Maxwell field for $v<0$ is given by (\ref{1.1}) and (\ref{1.4}) and if the space--time and 
Maxwell field for $u<0$ is given by (\ref{1.9}) and (\ref{1.10}) (the space--time for $v<0$ \emph{and} $u<0$ is trivially flat Minkowskian space--time with vanishing Maxwell field) then what is the post--collision 
space--time for $u>0, v>0$ and post--collision Maxwell field for $u>0, v>0$? The post--collision space--time and Maxwell field constitute the Bell--Szekeres solution of the Einstein--Maxwell vacuum field equations.  We give 
a derivation of this solution in section 2 which acts as a guide for solving the analogous problem for colliding gravitational shock waves in section 3.

The space--time model of the gravitational field of gravitational shock waves is given by the line element 
\begin{equation}\label{1.11}
ds^2=-\cos^2au_+dx^2-\cosh^2au_+dy^2+2\,du\,dv\ ,\end{equation}where $a$ is a real constant. This takes the form (\ref{1.2}) with now the basis 1--forms given by
\begin{equation}\label{1.12}
\vartheta^1=\cos au_+\,dx\ ,\ \ \vartheta^2=\cosh au_+\,dy\ ,\ \ \vartheta^3=dv\ ,\ \ \vartheta ^4=du\ .\end{equation}The components $R_{ab}$ of the Ricci tensor on the tetrad defined by the basis 1--forms vanish, so 
that the metric tensor given via the line element (\ref{1.11}) is a solution of Einstein's vacuum field equations. The corresponding Riemann curvature tensor components in Newman--Penrose notation are 
given by 
\begin{equation}\label{1.13}
\Psi_0=\Psi_1=\Psi_2=\Psi_3=0\ \ \ {\rm and}\ \ \ \Psi_4=a^2\vartheta(u)\ .\end{equation}This is a Petrov type N curvature tensor and thus represents pure gravitational radiation with propagation direction 
in space--time the degenerate principal null direction of the Riemann tensor $\partial/\partial v$. We have again here plane, shock waves which are homogeneous and have a step function profile, but in this case 
the waves are gravitational waves. For similar waves traveling in the opposite direction
\begin{equation}\label{1.14}
ds^2=-\cos^2bv_+dx^2-\cosh^2bv_+dy^2+2\,du\,dv\ ,\end{equation}where $b$ is a real constant. The analogue of the Bell and Szekeres problem is to ask: if the space--time for $v<0$ is given by (\ref{1.11}) 
and if the space--time for $u<0$ is given by (\ref{1.14}) (the space--time for $v<0$ \emph{and} $u<0$ is trivially flat Minkowskian space--time) then what is the post--collision 
space--time for $u>0, v>0$? Following the pattern of the Bell--Szekeres derivation in section 2 below we give the answer to this question in section 3. This is the main result of this paper. The most significant 
feature of this answer is that the appropriate field equations in the post--collision region of the space--time are Einstein's vacuum field equations with a cosmological constant. The cosmological constant, which could conceivably represent dark energy, is proportional to the product of the constants $a$ and $b$ associated with the incoming gravitational shock waves indicated above. Some important properties of the derivation 
are discussed in section 4 including a summary of the principal physical attributes of the solution derived in section 3.

\section{Derivation of Bell-Szekeres Solution}
\setcounter{equation}0\noindent
For the head--on collision of homogeneous plane waves, the post--collision region $u>0, v>0$ of space--time is well known to take the Rosen--Szekeres form \cite{Rosen}, \cite{S}, \cite{KP}
\begin{equation}\label{3.1}
ds^2=-e^{-U}(e^Vdx^2+e^{-V}dy^2)+2e^{-M}du\,dv\ ,\end{equation}where $U, V, M$ are each functions of $u, v$. For the case involving electromagnetic waves, which interests us particularly in this 
section, the Maxwell field has in general only two Newman-Penrose 
components $\Phi_0(u, v)$ and $\Phi_2(u, v)$, or equivalently, the Maxwell field is given by a 2--form
\begin{equation}\label{3.1'}
F=\frac{1}{2}F_{ab}\vartheta^a\wedge\vartheta^b=\Phi_0\vartheta^3\wedge\vartheta^1+\Phi_2\vartheta^1\wedge\vartheta^4\ ,\end{equation}which also serves to define $\Phi_0$ and $\Phi_2$ 
in terms of $F_{ab}$ and the basis 1--forms. It is convenient to write
\begin{equation}\label{3.1''}
\Phi_0=e^{M/2}\phi_0\ \ \ {\rm and}\ \ \ \Phi_2=e^{M/2}\phi_2\ ,\end{equation}and then Maxwell's equations in terms of $\phi_0(u, v)$ and $\phi_2(u, v)$ read:
\begin{eqnarray}
\frac{\partial\phi_0}{\partial u}&=&\frac{1}{2}U_u\,\phi_0-\frac{1}{2}V_v\,\phi_2\ ,\label{3.2}\\
\frac{\partial\phi_2}{\partial v}&=&\frac{1}{2}U_v\,\phi_2-\frac{1}{2}V_u\,\phi_0\ ,\label{3.3}\end{eqnarray}with the subscripts denoting partial differentiation, and the Einstein--Maxwell 
field equations read:
\begin{eqnarray}
U_{uv}&=&U_u\,U_v\ ,\label{3.4}\\
2U_{uu}&=&U_u^2+V_u^2-2U_uM_u+4\phi_2^2\ ,\label{3.5}\\
2U_{vv}&=&U_v^2+V_v^2-2M_vU_v+4\phi_0^2\ ,\label{3.6}\\
2V_{uv}&=&U_uV_v+U_vV_u+4\phi_0\phi_2\ ,\label{3.7}\\
2M_{uv}&=&V_uV_v-U_uU_v\ .\label{3.8}\end{eqnarray}We wish to solve these equations in the region $u>0, v>0$ with the following boundary conditions (which are obtained from (\ref{1.1}), (\ref{1.4}) and 
from (\ref{1.9}), (\ref{1.10}) respectively) :
\begin{equation}\label{3.9}
{\rm When}\  v=0:\ e^{-U}=\cos^2au\ ,\ \ V=0\ ,\ \ M=0\ ,\ \ \phi_2=a\ ;\end{equation}
\begin{equation}\label{3.10}
{\rm When}\  u=0:\ e^{-U}=\cos^2bv\ ,\ \ V=0\ ,\ \ M=0\ ,\ \ \phi_0=b\ ,\end{equation}where $a, b$ are constants. We note the well known fact that under a change of coordinates $u\rightarrow\bar u=\bar u(u)$ 
and $v\rightarrow\bar v=\bar v(v)$ the line element and field equations remain invariant in form with $M\rightarrow\bar M,\ \phi_0\rightarrow\bar\phi_0,\ \phi_2\rightarrow\bar\phi_2$ with
\begin{equation}\label{3.11}
e^{\bar M}=\frac{d\bar u}{du}\,\frac{d\bar v}{dv}e^M\ ,\ \ \bar\phi_0=\left (\frac{d\bar v}{dv}\right )^{-1}\phi_0\ ,\ \ \bar\phi_2=\left (\frac{d\bar u}{du}\right )^{-1}\phi_2\ .\end{equation}Eq.(\ref{3.4}) is 
equivalent to $(e^{-U})_{uv}=0$ and solving this with
\begin{equation}\label{3.12}
e^{-U}=-1+\cos^2au+\cos^2bv=\cos(au-bv)\,\cos(au+bv)\ ,\end{equation}satisfies the boundary conditions. 

We see from (\ref{3.10}) that when $u=0$ we have $V_v=0$ but we shall require $V_u$ and $\phi_2$ when $u=0$ which we write as $(V_u)_{u=0}$ and $(\phi_2)_{u=0}$. Also when $v=0$ 
we have $V_u=0$ and we shall require $(V_v)_{v=0}$ and $(\phi_0)_{v=0}$. To derive $(V_u)_{u=0}$ and $(\phi_2)_{u=0}$ we use (\ref{3.3}) and (\ref{3.7}) evaluated at $u=0$. First we note 
from (\ref{3.12}) that
\begin{eqnarray}
U_u&=&a\tan(au-bv)+a\tan(au+bv)\ ,\label{3.13}\\
U_v&=&-b\tan(au-bv)+b\tan(au+bv)\ ,\label{3.14}\end{eqnarray}and so $(U_u)_{u=0}=0$ and $(U_v)_{u=0}=2b\tan bv$. Hence (\ref{3.3}) and (\ref{3.7}) evaluated at $u=0$ yield
\begin{eqnarray}
\frac{d}{dv}\left\{(\phi_2)_{u=0}\cos bv\right\}&=&-\frac{b}{2}(V_u)_{u=0}\cos bv\ ,\label{3.15}\\
\frac{d}{dv}\left\{(V_u)_{u=0}\cos bv\right\}&=&2b\,(\phi_2)_{u=0}\cos bv\ ,\label{3.16}\end{eqnarray}respectively. Solving these we arrive at
\begin{equation}\label{3.17}
(\phi_2)_{u=0}=P\,\tan bv+Q\ \ \ {\rm and}\ \ \ (V_u)_{u=0}=-2P+2Q\,\tan bv\ ,\end{equation}where $P, Q$ are constants. When $u=0$ \emph{and} $v=0$ we have $\phi_2=a$ and $V_u=0$ and 
this implies that $P=0$ and $Q=a$. Therefore
\begin{equation}\label{3.18}
(\phi_2)_{u=0}=a\ \ {\rm and}\ \ (V_u)_{u=0}=2a\tan bv\ .\end{equation}Similarly using (\ref{3.2}) and (\ref{3.7}) evaluated at $v=0$ leads to 
\begin{equation}\label{3.19}
(\phi_0)_{v=0}=b\ \ {\rm and}\ \ (V_v)_{v=0}=2b\tan au\ .\end{equation}Now Maxwell's equations can be written in the form
\begin{equation}\label{3.20}
\frac{\partial}{\partial u}(\log\phi_0)=\frac{1}{2}U_u-\frac{1}{2}V_v\,\frac{\phi_2}{\phi_0}\ \ {\rm and}\ \ \frac{\partial}{\partial v}(\log\phi_2)=\frac{1}{2}U_v-\frac{1}{2}V_u\,\frac{\phi_0}{\phi_2}\ ,\end{equation}
from which we easily deduce that
\begin{equation}\label{3.21}
2\frac{\partial^2}{\partial u\partial v}\left (\log\frac{\phi_0}{\phi_2}\right )=\frac{\partial}{\partial u}\left (V_u\,\frac{\phi_0}{\phi_2}\right )-\frac{\partial}{\partial v}\left (V_v\,\frac{\phi_2}{\phi_0}\right )\ .\end{equation}
We now make \emph{the key assumption} that
\begin{equation}\label{3.22}
\frac{\phi_0}{\phi_2}=\frac{A(u)}{B(v)}\ ,\end{equation}for some functions $A(u), B(v)$. On account of (\ref{3.11}) this means that there exists a frame of reference $\bar u, \bar v$ such that $\bar\phi_0=\bar\phi_2$. 
This implies that $\bar\phi_0^2=\bar\phi_2^2$ and so if $\bar\phi_0, \bar\phi_2$ are thought of as describing two families of backscattered electromagnetic radiation following the collision, then we 
are making the simplifying assumption that in the barred 
frame the energy densities of the two families of backscattered waves are equal. From (\ref{3.18}) and (\ref{3.19}) it follows that (\ref{3.22}) implies
\begin{equation}\label{3.23}
\frac{\phi_0}{\phi_2}=\frac{b}{a}\ ,\end{equation}for $u\geq 0, v\geq 0$. Using this in (\ref{3.21}) we arrive at the wave equation
\begin{equation}\label{3.24}
\frac{1}{a}\frac{\partial}{\partial u}\left (\frac{1}{a}V_u\right )=\frac{1}{b}\frac{\partial}{\partial v}\left (\frac{1}{b}V_v\right )\ ,\end{equation}or, with $\bar u=au, \bar v=bv$,
\begin{equation}\label{3.25}
V_{\bar u\bar u}=V_{\bar v\bar v}\ .\end{equation}Solving this for $V(\bar u, \bar v)$ using the d'Alembert formula, with 
\begin{equation}\label{3.26}
V(0, \bar v)=0\ \ \ {\rm and}\ \ \ V_{\bar u}(0, \bar v)=2\tan\bar v\ ,\end{equation}by (\ref{3.10}) and (\ref{3.18}), we find that
\begin{equation}\label{3.27}
V(\bar u, \bar v)=\frac{1}{2}\int_{\bar v-\bar u}^{\bar v+\bar u}2\tan\xi\,d\xi=\log\frac{\cos(\bar u-\bar v)}{\cos(\bar u+\bar v)}\ ,\end{equation}and thus
\begin{equation}\label{3.28}
V(u, v)=\log\frac{\cos(au-bv)}{\cos(au+bv)}\ ,\end{equation}for $u\geq 0, v\geq 0$. Now
\begin{eqnarray}
V_u&=&-a\tan(au-bv)+a\tan(au+bv)\ ,\label{3.29}\\
V_v&=&b\tan(au-bv)+b\tan(au+bv)\ ,\label{3.30}\end{eqnarray}and so with (\ref{3.13}) and (\ref{3.14}) we find that $U_uU_v=V_uV_v$. Now from (\ref{3.8}) we have $M_{uv}=0$ 
from which it follows, using the boundary conditions (\ref{3.9}) and (\ref{3.10}), that $M=0$ for $u\geq 0, v\geq 0$. Substituting $U$ from (\ref{3.12}) and $V$ from (\ref{3.28}) into 
(\ref{3.5}) and (\ref{3.6}) with $M=0$ results in 
\begin{equation}\label{3.31}
\phi_0=b\ \ \ {\rm and}\ \ \ \phi_2=a\ ,\end{equation}for $u\geq 0, v\geq 0$, and now (\ref{3.7}) is automatically satisfied, as are the Maxwell equations (\ref{3.2}) and (\ref{3.3}) since 
$bU_u-aV_v=0=aU_v-bV_u$ on account of (\ref{3.13}), (\ref{3.14}), (\ref{3.29}) and (\ref{3.30}). The line element for $u>0, v>0$ is now
\begin{equation}\label{3.32}
ds^2=-\cos^2(au-bv)\,dx^2-\cos^2(au+bv)\,dy^2+2\,du\,dv\ .\end{equation}Equations (\ref{3.31}) and (\ref{3.32}) constitute the Bell--Szekeres solution of the Einstein--Maxwell vacuum field equations.

We can write the line element of the space--time in a unified form incorporating each of the four regions of the space--time I($u<0, v<0$), II($u>0, v<0$), III($u<0, v>0$) and IV($u>0, v>0$) as
\begin{equation}\label{3.33}
ds^2=-\cos^2(au_+-bv_+)\,dx^2-\cos^2(au_++bv_+)\,dy^2+2\,du\,dv\ .\end{equation}On the tetrad defined by the basis 1--forms $\vartheta ^1=\cos (au_+-bv_+)\,dx,\ 
\vartheta ^2=\cos (au_++bv_+)\,dy,\ \vartheta ^3=dv$ and $\vartheta^4=du$ the Ricci tensor components are 
\begin{eqnarray}
R_{ab}&=&-2\,ab\,\vartheta(u)\,\vartheta(v)\,\delta ^1_a\,\delta ^1_b+2\,ab\,\vartheta(u)\,\vartheta(v)\,\delta ^2_a\,\delta ^2_b\nonumber\\
&&-2\,b^2\vartheta(v)\delta ^3_a\,\delta ^3_b-2\,a^2\vartheta(u)\delta ^4_a\,\delta ^4_b\nonumber\\
&=&2\,E_{ab}\ ,\label{3.34}\end{eqnarray}where $E_{ab}$ is the electromagnetic energy--momentum tensor calculated with the Maxwell field
\begin{equation}\label{3.35}
F=a\,\vartheta(u)\vartheta ^1\wedge\vartheta^4+b\,\vartheta(v)\vartheta^3\wedge\vartheta^1\ .\end{equation}Finally the Newman--Penrose components of the Weyl conformal curvature tensor for the Bell--Szekeres space--time are given by
\begin{eqnarray}
\Psi_0&=&-b\,\delta (v)\,\tan au_+\ ,\label{3.36}\\
\Psi_1&=&0\ ,\label{3.37}\\
\Psi_2&=&0\ ,\label{3.38}\\
\Psi_3&=&0\ ,\label{3.39}\\
\Psi_4&=&-a\,\delta(u)\,\tan bv_+\ .\label{3.40}\end{eqnarray}For $u>0, v>0$ this space--time is conformally flat. The appearance of the Dirac delta functions in two of these components means 
that there is an impulsive gravitational wave with history $v=0, u>0$ described by $\Psi_0$ and an impulsive gravitational wave with history $u=0, v>0$ described by $\Psi_4$. Thus 
the energy in the incoming electromagnetic shock waves is re-distributed following the collision into two impulsive gravitational waves followed by the superposition of two systems of electromagnetic 
shock waves described by (\ref{3.35}).

\section{Collision of Gravitational Shock Waves}
\setcounter{equation}0\noindent
Taking the Bell--Szekeres example as our guide we consider now the head--on collision of the homogeneous, plane gravitational waves described by (\ref{1.11}) and (\ref{1.14}). 
Again for the post collision space--time we work with a line element of the form (\ref{3.1}) but now we are interested in Einstein's field equations with a cosmological constant $\Lambda$ given by $R_{ab}=\Lambda\,g_{ab}$. 
Written out explicitly these equations read:
\begin{eqnarray}
U_{uv}&=&U_u\,U_v-\Lambda\,e^{-M}\ ,\label{4.1}\\
2V_{uv}&=&U_uV_v+U_vV_u\ ,\label{4.2}\\
2U_{uu}&=&U_u^2+V_u^2-2M_uU_u\ ,\label{4.3}\\
2U_{vv}&=&U_v^2+V_v^2-2M_vU_v\ ,\label{4.4}\\
2M_{uv}&=&V_uV_v-U_uU_v\ .\label{4.5}\end{eqnarray}The Newman--Penrose components of the Weyl conformal curvature tensor are:
\begin{eqnarray}
\Psi_0&=&-\frac{1}{2}e^M\{V_{vv}+(M_v-U_v)V_v\}\ ,\label{4.6}\\
\Psi_1&=&0\ ,\label{4.7}\\
\Psi_2&=&\frac{1}{4}e^M(V_uV_v-U_uU_v)+\frac{1}{6}\Lambda\ ,\label{4.8}\\
\Psi_3&=&0\ ,\label{4.9}\\
\Psi_4&=&-\frac{1}{2}e^M\{V_{uu}+(M_u-U_u)V_u\}\ .\label{4.10}\end{eqnarray}The Bianchi identities are:
\begin{eqnarray}
\frac{\partial\Psi_0}{\partial u}&=&\left (M_u+\frac{1}{2}U_u\right )\Psi_0-\frac{3}{2}V_v\Psi_2\ ,\label{4.11}\\
\frac{\partial\Psi_4}{\partial v}&=&\left (M_v+\frac{1}{2}U_v\right )\Psi_4-\frac{3}{2}V_u\Psi_2\ ,\label{4.12}\\
\frac{\partial\Psi_2}{\partial u}&=&\frac{3}{2}U_u\Psi_2-\frac{1}{2}V_v\Psi_4\ ,\label{4.13}\\
\frac{\partial\Psi_2}{\partial v}&=&\frac{3}{2}U_v\Psi_2-\frac{1}{2}V_u\Psi_0\ .\label{4.14}\end{eqnarray}These can be calculated using the formulas given by Chandrasekhar \cite{C} or, 
alternatively, they can be calculated directly by taking the appropriate partial derivatives of (\ref{4.6})--(\ref{4.10}) and simplifying the results using the field equations (\ref{4.1})--(\ref{4.5}) and 
using again (\ref{4.6})--(\ref{4.10}). 
The boundary conditions in this case are:
\begin{equation}\label{4.15}
\noindent{\rm When}\  v=0:\ e^{-U}=\cos au\cosh au\ ,\ \ e^V=\frac{\cos au}{\cosh au}\ ,\ \ M=0\ ,\ \ \Psi_4=a^2\ ;\end{equation}
\begin{equation}\label{4.16}
\noindent{\rm When}\  u=0:\ e^{-U}=\cos bv\cosh bv\ ,\ \ e^V=\frac{\cos bv}{\cosh bv}\ ,\ \ M=0\ ,\ \ \Psi_0=b^2\ ,\end{equation}where $a, b$ are constants.

To begin with we shall require $(U_v)_{v=0}, (V_v)_{v=0}, (M_v)_{v=0}$ and $(\Psi_0)_{v=0}$. It is useful to note from the boundary conditions that
\begin{eqnarray}
(U_u)_{v=0}&=&a\tan au-a\tanh au\ ,\label{4.17}\\
(V_u)_{v=0}&=&-a\tan au-a\tanh au\ .\label{4.18}\end{eqnarray}Evaluating (\ref{4.1}) at $v=0$ results in
\begin{equation}\label{4.19}
\frac{d}{du}(U_v)_{v=0}+a(-\tan au+\tanh au)(U_v)_{v=0}=-\Lambda\ ,\end{equation}which we can rewrite as
\begin{equation}\label{4.20}
\frac{d}{du}\left (\cos au\cosh au(U_v)_{v=0}\right )=-\Lambda\cos au\cosh au\ ,\end{equation}and solve with 
\begin{equation}\label{4.21}
\cos au\cosh au(U_v)_{v=0}=-\frac{\Lambda}{2a}(\sin au\cosh au+\cos au\sinh au)+C\ ,\end{equation}where $C$ is a constant of integration. Since $U_v$ at $u=0$ \emph{and} 
$v=0$ vanishes we have $C=0$ and thus
\begin{equation}\label{4.22}
(U_v)_{v=0}=-\frac{\Lambda}{2a}(\tan au+\tanh au)\ .\end{equation}Similarly by evaluating (\ref{4.1}) at $u=0$ we obtain
\begin{equation}\label{4.23}
(U_u)_{u=0}=-\frac{\Lambda}{2b}(\tan bv+\tanh bv)\ .\end{equation}To determine $(V_v)_{v=0}$ we evaluate (\ref{4.2}) at $v=0$ and arrive at
\begin{equation}\label{4.24}
\frac{d}{du}(V_v)_{v=0}=\frac{a}{2}(\tan au-\tanh au)(V_v)_{v=0}+\frac{\Lambda}{4}(\tan au+\tanh au)^2\ .\end{equation}To solve this we make use of the identity
\begin{equation}\label{4.25}
\frac{\Lambda}{4}(\tan au+\tanh au)^2=\frac{\Lambda}{2a}\frac{d}{du}(\tan au-\tanh au)-\frac{\Lambda}{4}(\tan au-\tanh au)^2\ .\end{equation}This enables us to write (\ref{4.24}) 
in the form
\begin{equation}\label{4.26}
\frac{dW}{du}-\frac{a}{2}(\tan au-\tanh au)W=0\ ,\end{equation}with 
\begin{equation}\label{4.27}
W=(V_v)_{v=0}-\frac{\Lambda}{2a}(\tan au-\tanh au)\ .\end{equation}Now (\ref{4.26}) reads
\begin{equation}\label{4.28}
\frac{d}{du}\{(\cos au\cosh au)^{1/2}W\}=0\ ,\end{equation}and thus we arrive at
\begin{equation}\label{4.29}
(V_v)_{v=0}=\frac{\Lambda}{2a}(\tan au-\tanh au)+C_1(\cos au\cosh au)^{-1/2}\ ,\end{equation}where $C_1$ is a constant of integration. Since $V_v$ vanishes when $u=0$ \emph{and} $v=0$ 
we have $C_1=0$ and thus
\begin{equation}\label{4.30}
(V_v)_{v=0}=\frac{\Lambda}{2a}(\tan au-\tanh au)\ .\end{equation}Proceeding similarly with (\ref{4.2}) evaluated at $u=0$ we find that
\begin{equation}\label{4.31}
(V_u)_{u=0}=\frac{\Lambda}{2b}(\tan bv-\tanh bv)\ .\end{equation}Next we require $(M_v)_{v=0}$ and this is obtained by evaluating (\ref{4.5}) at $v=0$. From (\ref{4.17}), (\ref{4.18}), 
(\ref{4.22}) and (\ref{4.30}) we see that the right hand side of (\ref{4.5}) evaluated at $v=0$ vanishes and so we have 
\begin{equation}\label{4.32}
\frac{d}{du}(M_v)_{v=0}=0\ ,\end{equation}and thus $(M_v)_{v=0}=C_2={\rm constant}$. But $M_v$ vanishes when $u=0$ \emph{and} $v=0$ and so $C_2=0$. Hence
\begin{equation}\label{4.33}
(M_v)_{v=0}=0\ .\end{equation}Evaluating (\ref{4.5}) at $u=0$ similarly results in 
\begin{equation}\label{4.34}
(M_u)_{u=0}=0\ .\end{equation}Finally we shall require $(\Psi_0)_{v=0}$. The expression for $\Psi_2$ given by (\ref{4.8}) evaluated at $v=0$, remembering (\ref{4.15}), yields
\begin{equation}\label{4.35}
(\Psi_2)_{v=0}=\frac{1}{6}\Lambda\ .\end{equation}Using this, along with (\ref{4.17}) and (\ref{4.30}), in the Bianchi identity (\ref{4.11}) evaluated at $v=0$ we find
\begin{equation}\label{4.36}
\frac{d}{du}(\Psi_0)_{v=0}-\frac{1}{2}a(\tan au-\tanh au)(\Psi_0)_{v=0}=-\frac{\Lambda^2}{8a}(\tan au-\tanh au)\ ,\end{equation}which can be rewritten in the form
\begin{equation}\label{4.37}
\frac{d}{du}\{(\cos au\cosh au)^{1/2}(\Psi _0)_{v=0}\}=\frac{\Lambda^2}{4a^2}\frac{d}{du}(\cos au\cosh au)^{1/2}\ ,\end{equation}which leads to 
\begin{equation}\label{4.38}
(\Psi_0)_{v=0}=\frac{\Lambda^2}{4a^2}+\frac{C_3}{(\cos au\cosh au)^{1/2}}\ ,\end{equation}where $C_3$ is a constant of integration. Now we can summarize the current situation 
with regard to the Weyl tensor evaluated at $v=0$ as follows:
\begin{equation}\label{4.39}
(\Psi_0)_{v=0}=\frac{\Lambda^2}{4a^2}+\frac{C_3}{(\cos au\cosh au)^{1/2}}\ ,\ \ (\Psi_2)_{v=0}=\frac{\Lambda}{6}\ ,\ \ (\Psi_4)_{v=0}=a^2\ .\end{equation}We will return to these equations after 
we have discussed, for the present case, the analogue of (\ref{3.21}) with (\ref{3.22}). 

From (\ref{4.11}) we have
\begin{equation}\label{4.40}
\frac{\partial}{\partial u}\log\Psi_0=M_u+\frac{1}{2}U_u-\frac{3}{2}V_v\frac{\Psi_2}{\Psi_0}\ ,\end{equation}and from (\ref{4.12}) 
\begin{equation}\label{4.41}
\frac{\partial}{\partial v}\log\Psi_4=M_v+\frac{1}{2}U_v-\frac{3}{2}V_u\frac{\Psi_2}{\Psi_4}\ ,\end{equation}from which we deduce that
\begin{equation}\label{4.42}
\frac{2}{3}\frac{\partial^2}{\partial u\partial v}\log\left(\frac{\Psi_4}{\Psi_0}\right )=\frac{\partial}{\partial v}\left (V_v\frac{\Psi_2}{\Psi_0}\right )-\frac{\partial}{\partial u}\left (V_u\frac{\Psi_2}{\Psi_4}\right )\ .\end{equation}
Next (\ref{4.13}) and (\ref{4.14}) can be written as
\begin{eqnarray}
\frac{\partial}{\partial u}\log\Psi_2&=&\frac{3}{2}U_u-\frac{1}{2}V_v\frac{\Psi_4}{\Psi_2}\ ,\label{4.43}\\
\frac{\partial}{\partial v}\log\Psi_2&=&\frac{3}{2}U_v-\frac{1}{2}V_u\frac{\Psi_0}{\Psi_2}\ ,\label{4.44}\end{eqnarray}and thus we have
\begin{equation}\label{4.45}
\frac{\partial}{\partial v}\left (V_v\frac{\Psi_4}{\Psi_2}\right )-\frac{\partial}{\partial u}\left (V_u\frac{\Psi_0}{\Psi_2}\right )=0\ .\end{equation}
We have here two equations, (\ref{4.42}) and (\ref{4.45}), where we had one equation (\ref{3.21}) in the electromagnetic case. If we make the assumption that 
\begin{equation}\label{4.46}
\Psi_2=k\Psi_0^{1/2}\Psi_4^{1/2}\ ,\end{equation}for some constant $k$ to be determined later, we can arrive at a situation exactly analogous to the electromagnetic case. With 
this assumption (\ref{4.42}) becomes
\begin{equation}\label{4.47}
\frac{2}{3k}\frac{\partial^2}{\partial u\partial v}\log\left(\frac{\Psi_4}{\Psi_0}\right )=\frac{\partial}{\partial v}\left (V_v\frac{\Psi_4^{1/2}}{\Psi_0^{1/2}}\right )
-\frac{\partial}{\partial u}\left (V_u\frac{\Psi_0^{1/2}}{\Psi_4^{1/2}}\right )\ ,\end{equation}while (\ref{4.45}) now reads
\begin{equation}\label{4.48}
\frac{\partial}{\partial v}\left (V_v\frac{\Psi_4^{1/2}}{\Psi_0^{1/2}}\right )
-\frac{\partial}{\partial u}\left (V_u\frac{\Psi_0^{1/2}}{\Psi_4^{1/2}}\right )=0\ .\end{equation}Thus we now have the analogue of the electromagnetic case in that 
\begin{equation}\label{4.49}
\frac{\Psi_4}{\Psi_0}=\frac{A(u)}{B(v)}\ ,\end{equation}for some functions $A(u)$ and $B(v)$ and then $V(u,v)$ is determined by (\ref{4.48}) (the analogue of (\ref{3.24})). 

Before proceeding it is useful to now 
compute the constant of integration $C_3$ in (\ref{4.38}). We see that when (\ref{4.46}) is evaluated at $v=0$ we must have $(\Psi_0)_{v=0}$ constant, since $(\Psi_2)_{v=0}$ and 
$(\Psi_4)_{v=0}$ are constants given in (\ref{4.39}). Therefore $C_3=0$ and so 
\begin{equation}\label{4.50}
(\Psi_0)_{v=0}=\frac{\Lambda^2}{4a^2}\ .\end{equation}But from (\ref{4.16}) we have $\Psi_0=b^2$ when $u=0$ \emph{and} $v=0$ and so $\Lambda^2=4a^2b^2$ and we can take 
\begin{equation}\label{4.51}
\Lambda=2ab\ ,\end{equation}from now on since, in particular, the signs of $a$ and $b$ are free to specify. This demonstrates the importance of including the cosmological constant in the field equations at the beginning of this derivation. Thus we have
\begin{equation}\label{4.52}
(\Psi_0)_{v=0}=b^2\ ,\ \ (\Psi_2)_{v=0}=\frac{1}{3}ab\ ,\ \ (\Psi_4)_{v=0}=a^2\ ,\end{equation}and so (\ref{4.46}) evaluated at $v=0$ yields $k=\frac{1}{3}$. 

We now have in particular $(\Psi_0)_{v=0}=b^2$ and $(\Psi_4)_{v=0}=a^2$ and similarly we find that $(\Psi_4)_{u=0}=a^2$ and $(\Psi_0)_{u=0}=b^2$. Hence it follows 
from (\ref{4.49}) that for $u\geq0, v\geq0$, 
\begin{equation}\label{4.53}
\frac{\Psi_4}{\Psi_0}=\frac{a^2}{b^2}\ .\end{equation}Consequently with $\bar u=au, \bar v=bv$ we see, following (\ref{4.48}), that $V(\bar u, \bar v)$ satisfies the wave equation
\begin{equation}\label{4.54}
V_{\bar v\bar v}=V_{\bar u\bar u}\ .\end{equation}Using the d'Alembert formula with (by (\ref{4.16}) and (\ref{4.31}))  
\begin{equation}\label{4.55}
V(0, \bar v)=\log\frac{\cos\bar v}{\cosh\bar v}\ ,\ \ {\rm and}\ \ \ V_{\bar u}(0, \bar v)=\tan\bar v-\tanh\bar v\ ,\end{equation}we have
\begin{equation}\label{4.56}
V(\bar u, \bar v)=\frac{1}{2}\log\left (\frac{\cos(\bar u+\bar v)\,\cos(\bar v-\bar u)}{\cosh(\bar u+\bar v)\,\cosh(\bar v-\bar u)}\right )+\frac{1}{2}\int_{\bar v-\bar u}^{\bar v+\bar u}\tan\xi-\tanh\xi\,d\xi\ ,\end{equation}
which simplifies to 
\begin{equation}\label{4.57}
V=\log\left (\frac{\cos(au-bv)}{\cosh(au+bv)}\right )\ ,\end{equation}for $u\geq0, v\geq0$.

We do not yet know $\Psi_0, \Psi_2, \Psi_4$ individually for $u>0, v>0$. However we do know the following relations between them which hold for $u\geq0, v\geq0$:
\begin{equation}\label{4.58}
\frac{\Psi_4}{\Psi_0}=\frac{a^2}{b^2}\ ,\qquad 3\Psi_2=\Psi_0^{1/2}\Psi_4^{1/2}\ ,\end{equation}and thus
\begin{equation}\label{4.59}
3\,\Psi_2=\frac{b}{a}\Psi_4=\frac{a}{b}\Psi_0\ .\end{equation}Using the latter in the Bianchi identities (\ref{4.11})--(\ref{4.14}) we have
\begin{eqnarray}
\frac{3\,b}{a}\frac{\partial\Psi_2}{\partial u}&=&\left (M_u+\frac{1}{2}U_u\right )\frac{3\,b}{a}\Psi_2-\frac{3}{2}V_v\Psi_2\ ,\label{4.60}\\
\frac{3\,a}{b}\frac{\partial\Psi_2}{\partial v}&=&\left (M_v+\frac{1}{2}U_v\right )\frac{3\,a}{b}\Psi_2-\frac{3}{2}V_u\Psi_2\ ,\label{4.61}\\
\frac{\partial\Psi_2}{\partial u}&=&\frac{3}{2}U_u\Psi_2-\frac{3\,a}{2\,b}V_v\Psi_2\ ,\label{4.62}\\
\frac{\partial\Psi_2}{\partial v}&=&\frac{3}{2}U_v\Psi_2-\frac{3\,b}{2\,a}V_u\Psi_2\ .\label{4.63}\end{eqnarray}Thus (\ref{4.60}) and (\ref{4.62}) with $\Psi_2\neq 0$ yield
\begin{equation}\label{4.64}
M_u-U_u=-\frac{a}{b}\,V_v\ ,\end{equation}and (\ref{4.61}) and (\ref{4.63}) with $\Psi_2\neq 0$ yield
\begin{equation}\label{4.65}
M_v-U_v=-\frac{b}{a}\,V_u\ .\end{equation}With $V$ given by (\ref{4.57}) the first of these two equations reads
\begin{eqnarray}
M_u-U_u&=&-a\tan(au-bv)+a\tanh(au+bv)\ ,\nonumber\\
&=&\frac{\partial}{\partial u}\left (\log\cos(au-bv)+\log\cosh(au+bv)\right )\ .\label{4.66}\end{eqnarray}Integrating and using the boundary conditions (\ref{4.16}) results in
\begin{equation}\label{4.67}
M-U=\log\left \{\cos(au-bv)\cosh(au+bv)\right \}\ .\end{equation}We easily see that this satisfies (\ref{4.65}) with again $V$ given by (\ref{4.57}). At this point we do not 
know $M$ and $U$ separately for $u>0, v>0$. Substituting for $U$ in terms of $M$ from (\ref{4.67}) and for $V$ from (\ref{4.57}) into the field equations (\ref{4.3}) and (\ref{4.4}) results 
in the equations
\begin{equation}\label{4.68}
2\,M_{uu}=-M_u^2\ ,\end{equation}and
\begin{equation}\label{4.69}
2\,M_{vv}=-M_v^2\ ,\end{equation}respectively. Since $M_u=0$ when $u=0$ and when $v=0$, and in addition $M_v=0$ when $u=0$ and when $v=0$, the only solutions of (\ref{4.68}) 
and (\ref{4.69}) which satisfy the boundary conditions are
\begin{equation}\label{4.70}
M_u=0=M_v\ ,\end{equation}for $u\geq0, v\geq0$ and thus it follows from the boundary conditions (\ref{4.15}) and (\ref{4.16}) that
\begin{equation}\label{4.71}
M=0\ ,\end{equation}for $u\geq0, v\geq0$. Hence we have from (\ref{4.67}):
\begin{equation}\label{4.72}
U=-\log\{\cos(au-bv)\cosh(au+bv)\}\ ,\end{equation}for $u\geq0, v\geq0$.

With $V, M, U$ given by (\ref{4.57}), (\ref{4.71}) and (\ref{4.72}) respectively the line element (\ref{3.1}) reads
\begin{equation}\label{4.73}
ds^2=-\cos^2(au-bv)\,dx^2-\cosh^2(au+bv)\,dy^2+2\,du\,dv\ ,\end{equation}
for $u\geq0, v\geq0$. Calculating $\Psi_0, \Psi_2, \Psi_4$ from (\ref{4.6}), (\ref{4.8}) and (\ref{4.10}), using $V, M, U$ given by (\ref{4.57}), (\ref{4.71}) and (\ref{4.72}),
results in 
\begin{equation}\label{4.74}
\Psi_0=b^2\ ,\ \ \Psi_2=\frac{1}{3}ab\ ,\ \ \Psi_4=a^2\ ,\end{equation}for $u\geq0, v\geq0$. It is now straightforward to check that (\ref{4.11})--(\ref{4.14}) are satisfied since 
$b\,U_u=a\,V_v$ and $a\,U_v=b\,V_u$ with $V$ given by (\ref{4.57}) and $U$ by (\ref{4.72}).

\section{Discussion}
\setcounter{equation}0\noindent
We have seen that an important part of the argument in section 3 is the assumption (\ref{4.46}) relating $\Psi_2$ to $\Psi_0$ and $\Psi_4$. It is clear that, as a consequence of the field 
equations (\ref{4.42}) and (\ref{4.45}), (\ref{4.46}) \emph{implies} the separation of variables (\ref{4.49}) in the function $\Psi_4/\Psi_0$. \emph{The converse is also true}. If the separation 
of variables (\ref{4.49}) is assumed then, as a consequence of the field equations (\ref{4.42}) and (\ref{4.45}), the equation (\ref{4.46}) holds. To see this we note that if the separation 
of variables (\ref{4.49}) is assumed then (\ref{4.42}) reads
\begin{equation}\label{5.1}
\frac{\partial}{\partial v}\left (V_v\frac{\Psi_2}{\Psi_0}\right )-\frac{\partial}{\partial u}\left (V_u\frac{\Psi_2}{\Psi_4}\right )=0\ .\end{equation}Putting
\begin{equation}\label{5.2}
\xi=\frac{\Psi_2}{\Psi_0^{1/2}\Psi_4^{1/2}}\ ,\end{equation}we can rewrite (\ref{5.1}) as
\begin{equation}\label{5.3}
\frac{\partial}{\partial v}\left (V_v\frac{\Psi_4^{1/2}}{\Psi_0^{1/2}}\xi\right )=\frac{\partial}{\partial u}\left (V_u\frac{\Psi_0^{1/2}}{\Psi_4^{1/2}}\xi\right )\ .\end{equation}Using (\ref{5.2}) again 
we can rewrite (\ref{4.45}) as
\begin{equation}\label{5.4}
\frac{\partial}{\partial v}\left (V_v\frac{\Psi_4^{1/2}}{\Psi_0^{1/2}}\xi^{-1}\right )=\frac{\partial}{\partial u}\left (V_u\frac{\Psi_0^{1/2}}{\Psi_4^{1/2}}\xi^{-1}\right )\ .\end{equation}Writing these 
equations out more explicitly we have 
\begin{equation}
\left (V_v\frac{\Psi_4^{1/2}}{\Psi_0^{1/2}}\right )\xi_v-\left (V_u\frac{\Psi_0^{1/2}}{\Psi_4^{1/2}}\right )\xi_u=-\left\{\frac{\partial}{\partial v}\left (V_v\frac{\Psi_4^{1/2}}{\Psi_0^{1/2}}\right )
-\frac{\partial}{\partial u}\left (V_u\frac{\Psi_0^{1/2}}{\Psi_4^{1/2}}\right )\right\}\xi\ ,\label{5.5}\end{equation}
\begin{equation}
-\left (V_v\frac{\Psi_4^{1/2}}{\Psi_0^{1/2}}\right )\xi_v+\left (V_u\frac{\Psi_0^{1/2}}{\Psi_4^{1/2}}\right )\xi_u=-\left\{\frac{\partial}{\partial v}\left (V_v\frac{\Psi_4^{1/2}}{\Psi_0^{1/2}}\right )
-\frac{\partial}{\partial u}\left (V_u\frac{\Psi_0^{1/2}}{\Psi_4^{1/2}}\right )\right\}\xi\ .\label{5.6}\end{equation}Adding (\ref{5.5}) and (\ref{5.6}) with $\xi\neq0$ results in (\ref{4.48}) and then 
(\ref{5.5}) and (\ref{5.6}) each become
\begin{equation}\label{5.7}
\left (V_v\frac{\Psi_4^{1/2}}{\Psi_0^{1/2}}\right )\xi_v-\left (V_u\frac{\Psi_0^{1/2}}{\Psi_4^{1/2}}\right )\xi_u=0\ .\end{equation}We now have before us the wave equation (\ref{4.48}) and this equation 
(\ref{5.7}). Under the coordinate transformation $u\rightarrow\bar u=\bar u(u)$ and $v\rightarrow\bar v=\bar v(v)$, mentioned following eq.(\ref{3.10}) above, $\Psi_0\rightarrow\bar\Psi_0=
\frac{d\bar u}{du}\left (\frac{d\bar v}{dv}\right )^{-1}\Psi_0,\ \Psi_2\rightarrow\bar\Psi_2=\Psi_2$ and $\Psi_4\rightarrow\bar\Psi_4=
\frac{d\bar v}{dv}\left (\frac{d\bar u}{du}\right )^{-1}\Psi_4$ and thus $\xi\rightarrow\bar\xi=\xi$. Hence if we choose 
\begin{equation}\label{5.8}
\frac{d\bar u}{du}=\left (A(u)\right )^{1/2}\ \ \ {\rm and}\ \ \ \frac{d\bar v}{dv}=\left (B(v)\right )^{1/2}\ ,\end{equation}with $A(u), B(v)$ appearing in the presently assumed separation of variables 
(\ref{4.49}) then $\bar\Psi_4=\bar\Psi_0$ and (\ref{5.7}) and (\ref{4.48}) simplify to read
\begin{equation}\label{5.9}
V_{\bar v}\,\xi_{\bar v}=V_{\bar u}\,\xi_{\bar u}\ ,\end{equation}and
\begin{equation}\label{5.10}
V_{\bar v\bar v}=V_{\bar u\bar u}\ ,\end{equation}respectively. The general solution of the wave equation (\ref{5.10}) is
\begin{equation}\label{5.11}
V(\bar u, \bar v)=f(\bar u+\bar v)+g(\bar u-\bar v)\ ,\end{equation}where the functions $f, g$ are arbitrary functions of their arguments. Substituting into (\ref{5.9}) we have 
\begin{equation}\label{5.12}
(\xi_{\bar u}-\xi_{\bar v})\,f'+(\xi_{\bar u}+\xi_{\bar v})\,g'=0\ ,\end{equation}where the primes on $f, g$ denote differentiation of these functions with respect to their arguments. If (\ref{5.12}) 
is to hold for $f, g$ arbitrary then we must have $\xi_{\bar u}=0=\xi_{\bar v}$ which implies that $\xi={\rm constant}$ and thus (\ref{5.2}) agrees with (\ref{4.46}).

As in section 2 we can write the line element of the space--time in section 3 in a unified form incorporating each of the four regions of the space--time I($u<0, v<0$), II($u>0, v<0$), III($u<0, v>0$) and IV($u>0, v>0$) as
\begin{equation}\label{5.13}
ds^2=-\cos^2(au_+-bv_+)\,dx^2-\cosh^2(au_++bv_+)\,dy^2+2\,du\,dv\ .\end{equation}On the tetrad defined by the basis 1--forms $\vartheta ^1=\cos (au_+-bv_+)\,dx,\ 
\vartheta ^2=\cosh (au_++bv_+)\,dy,\ \vartheta ^3=dv$ and $\vartheta^4=du$ the Ricci tensor components are 
\begin{eqnarray}
R_{ab}&=&b\,\delta (v)\,(\tan au_++\tanh au_+)\,\delta ^3_a\,\delta ^3_b+a\,\delta (u)\,(\tan bv_++\tanh bv_+)\,\delta ^4_a\,\delta ^4_b\nonumber\\
&&+2\,ab\,\vartheta(u)\,\vartheta(v)\,g_{ab}\ .\label{5.14}\end{eqnarray}The Weyl conformal curvature tensor components in Newman--Penrose notation are
\begin{eqnarray}
\Psi_0&=&b^2\vartheta(v)+\frac{1}{2}b\,(\tanh au_+-\tan au_+)\,\delta (v)\ ,\label{5.15}\\
\Psi_1&=&0\ ,\label{5.16}\\
\Psi_2&=&\frac{1}{3}ab\,\vartheta(u)\,\vartheta(v)\ ,\label{5.17}\\
\Psi_3&=&0\ ,\label{5.18}\\
\Psi_4&=&a^2\vartheta(u)+\frac{1}{2}a\,(\tanh bv_+-\tan bv_+)\,\delta (u)\ .\label{5.19}\end{eqnarray}For $u>0, v>0$ this Weyl tensor is Type D in the 
Petrov classification. The situation following the collision of the two gravitational shock waves is more complicated than in the electromagnetic case. Now the energy in the incoming 
waves is re-distributed into (1) two light--like shells (see \cite{BH}), which could correspond to bursts of neutrinos, described by the Dirac delta function terms in the Ricci tensor components (\ref{5.14}), 
(2) two impulsive gravitational waves described by the Dirac delta function terms in the Weyl tensor components (\ref{5.15}) and (\ref{5.19}) and (3) dark energy in the region $u>0, v>0$ described 
by the Ricci tensor in this region of space--time having the form
\begin{equation}\label{5.20}
R_{ab}=\Lambda\,g_{ab}\ \ \ {\rm with}\ \ \ \Lambda=2\,a\,b\ .\end{equation}

It is well known that the Bell--Szekeres space--time with line element (\ref{3.32}) can be written as the sum of the line elements of two 2--dimensional manifolds of equal constant curvature \cite{Mac}. This 
is the so--called Bertotti--Robinson space--time \cite{B}, \cite{R}. Motivated by this we have derived the line element (\ref{4.73}) as the sum of the line elements of two 
2--dimensional manifolds of constant curvature differing only in sign (see \cite{BH2}, \cite{BH3}). This is the so--called Nariai--Bertotti space--time \cite{B}, \cite{N}. A particularly simple 
example of a collision of plane, homogeneous, light--like signals combining gravitational and electromagnetic shock waves in each in--coming signal has recently been given in \cite{BH4}.


\begin{thebibliography}{99}
\bibitem{Pe} P. J. E. Peebles and B. Ratra, Rev. Mod. Phys. {\bf 75}, 559 (2003).
\bibitem{BS} P. Bell and P. Szekeres, Gen. Rel. Grav. {\bf 5}, 275 (1974).
\bibitem{NP} E. T. Newman and R. Penrose, J. Math. Phys. {\bf 3}, 566 (1962).
\bibitem{Rosen} N. Rosen, Phys. Z. Sowjet {\bf 12}, 366 (1937).
\bibitem{S} P. Szekeres, J. Math. Phys. {\bf 13}, 286 (1972).
\bibitem{KP} K. A. Khan and R. Penrose, Nature {\bf 229}, 185 (1971).
\bibitem{C} S. Chandrasekhar, \emph{The Mathematical Theory of Black Holes}, Oxford Classic Texts in the Physical Sciences, (Clarendon Press, Oxford 1998).
\bibitem{BH} C. Barrab\`es and P. A. Hogan, \emph{Singular Null Hypersurfaces in General Relativity} (World Scientific, Singapore, 2004).
\bibitem{Mac} H. Stephani, D. Kramer, M. A. H. MacCallum, C. Hoenselaars  and E. Herlt, \emph{Exact Solutions of Einstein's Equations}, 2nd. edition (Cambridge University Press, Cambridge 2003). 
\bibitem{B} B. Bertotti, Phys. Rev. {\bf 116}, 1331 (1959).
\bibitem{R} I. Robinson, Bull. Acad. Polon. {\bf 7}, 351 (1959).
\bibitem{BH2} C. Barrab\`es and P. A. Hogan, Progress of Theor. Physics {\bf 126}, 1 (2011).
\bibitem{BH3} C. Barrab\`es and P. A. Hogan, \emph{Advanced General Relativity: Gravity Waves, Spinning Particles and Black Holes}, International Series of Monographs on Physics no.160 (Oxford 
University Press, Oxford 2013).
\bibitem{N} H. Nariai, Gen. Rel. Grav. {\bf 31}, 963 (1999), reprinted from Reports of Tohoku University (1951).
\bibitem{BH4} C. Barrab\`es and P. A. Hogan, Phys. Rev. D{\bf 88}, 087501 (2013).
\end{thebibliography}
\end{document}